\documentclass[11pt,a4paper]{article}

\usepackage{geometry}
\usepackage{amsfonts}
\usepackage{amsmath}
\usepackage{amssymb}

\usepackage{mathrsfs}
\usepackage{wasysym}

\usepackage{bbm}
\usepackage{bm}
\usepackage{graphicx}
\usepackage{hyperref}

\usepackage{authblk}

\newcommand{\sfrac}{\textstyle \frac}
\newcommand{\st}[1]{\text{\tiny \rm #1}}

\def\bq{\begin{equation}}
\def\ee{\end{equation}}

\newtheorem{Theorem}{Theorem}[section]

\renewcommand{\d}{{\rm d}}

\title{\huge Total Collisions in the N-Body Shape Space}

  \author[a]{Flavio Mercati\thanks{fmercati@ubu.es or flavio.mercati@gmail.com.}}
    \author[b]{Paula Reichert\thanks{reichert@math.lmu.de or paula.reichert@gmail.com.}}
          
\affil[a]{\emph{Departamento de F\'isica, Universidad de Burgos,
 09001 Burgos, Spain.}}\vspace{0.3cm}

\affil[b]{\emph{Mathematisches Institut, Ludwig-Maximilians-Universit\"at
 M\"unchen, Germany.}}

\date{}

\begin{document}
\maketitle
\thispagestyle{empty}

\begin{abstract}

\noindent We discuss the total collision singularities of the gravitational $N$-body problem on shape space. Shape space is the relational configuration space of the system obtained by quotienting ordinary configuration space with respect to the similarity group of total translations, rotations, and scalings. For the zero-energy gravitating $N$-body system, the dynamics on shape space can be constructed explicitly and the points of total collision, which are the points of central configuration and zero shape momenta, can be analyzed in detail. It turns out that, even on shape space where scale is not part of the description, the equations of motion diverge at (and only at) the points of total collision. We construct and study the stratified total-collision manifold and show that, at the points of total collision on shape space, the singularity is essential. There is, thus, no way to evolve solutions through these points. This mirrors closely the big bang singularity of general relativity, where the homogeneous-but-not-isotropic cosmological model of Bianchi IX shows an essential singularity at the big bang. A simple modification of the general-relativistic model (the addition of a stiff matter field) changes the system into one whose shape-dynamical description allows for a deterministic evolution through the singularity. We suspect that, similarly, some modification of the dynamics would be required in order to regularize the total collision singularity of the \emph{N}-body model.\end{abstract}

\newpage

\section{Introduction}
\pagenumbering{arabic}

Ever since Newton published the \emph{Principia} in 1687, the~gravitational $N$-body problem and, with~it, the~total collision singularity has been an object of intensive study among mathematicians, starting from Euler and Lagrange who found special solutions to the Newtonian 3-body problem, up~to Poincar\'e who famously received the price of the King of Sweden for his proposal of a general solution to it---a work he had to withdraw due to errors (still, it was a brilliant work and, in~a revised form, became the foundations of chaos~theory).  

It was then Sundman who solved the 3-body problem in 1907~\cite{Sundman1907,Sundman1912}. Sundman made use of the fact---which he proved---that total collisions can occur only if the total angular momentum $\bm L$ of the system is zero ($\bm L=0$). By~applying a convenient regularization procedure (change of variables) and using the fact that, for~$N=3$, all $\bm L \neq 0$ solutions can be bounded away from the triple collision, Sundman was able to provide a general $\bm L \neq 0$ solution to the 3-body problem in form of a convergent infinite power~series.

The  problem of total collisions in the 3-body model was the subject of a number of studies~\cite{Siegel967,Simo1980,Easton1971, McGehee1974}, which concluded that the total collisions cannot be regularized unless the masses take some exceptional values. Still, the~problem of $N$-body collisions remained untouched and while there exists a proof for the analytic continuation of solutions through binary collisions, nothing of that kind exists for $N\ge 3$ (cf. Saari~\cite{Saari2005}). 

In this paper, we discuss total collisions on shape space. Shape space is the relational configuration space of the system which is obtained from ordinary configuration space by quotienting with respect to the similarity group of translations, rotations, and~scalings (dilations). It has been shown in~\cite{Barbour:2013} (see also~\cite{FlavioSDBook}) that there exists a unique description of the $E\ge 0$ Newtonian $N$-body system on shape space (where $E$ refers to the total energy of the system).

Now, since on shape space, scale is no longer part of the description, one might hope to pass the singularity of a total collision by (uniquely) evolving the shape degrees of freedom through that point. If~that were possible, one could connect two total-collision solutions from absolute space, one with a collision in its past, one with a collision in its future, to~form \emph{one} solution passing the point of $N$-body collision (the Big Bang of the $E\ge 0$ Newtonian universe).

Unfortunately, this is not the case. Although there exists a unique description of total collisions on shape space---which is interesting in itself since this description is purely shape-dynamical, i.e.,~free of scale---the shape dynamics turns out to be singular precisely at (and only at) these points. Even more, one finds that the singularity is in genera essential, unless~the ratios between the particle masses take special~values.

In this paper, we explicitly analyze the way in which solutions run into the singularity on shape space and construct the stratified manifold of total-collision solutions. This will constitute Section~\ref{sec3}, the~main section of this paper. Section~\ref{sec2} will contain an overview of Chazy's noteworthy 1918 result~\cite{Chazy1918} on the asymptotic behavior of solutions at the points of total collision (a result which has been rediscovered much later by Saari~\cite{Saari1984} and which we use to identify total collisions on shape space). Section~\ref{sec4} will finally compare the result we have obtained for the $N$-body system to the general-relativistic Bianchi IX model, where the shape-dynamical description allows for a continuation of solutions through the point of zero volume (the Big Bang).

\section{Chazy's 1918 Proof of the Total Collision~Theorem}\label{sec2}

Consider the gravitational $N$-body problem, involving $N$ point particles of mass $m_a$ with coordinates $\bm r_a \in \mathbbm{R}^3$ and momenta $\bm p^a \in \mathbbm{R}^3$, $a= 1 ,\dots, N$, and~Hamiltonian
\begin{equation}
H_\st{New} = \sum_{a=1}^N \frac{\| \bm p_a \|^2}{2 m_a} + V_\st{New} \,, \qquad  V_\st{New} = - \sum_{a<b} \frac{m_a m_b}{\| \bm r_a - \bm r_b \| } \,.
\end{equation}

Chazy~\cite{Chazy1918} was the first to prove the following theorem:
\begin{Theorem}
A total collision  ($\bm r_a = \bm r_b ~~ \forall ~a,b$) can only happen if the total angular momentum $\bm L = \sum_{a=1}^N \bm  r_a \times \bm p_a$ is zero and at a central configuration, that is, a~configuration such that
\begin{equation}\label{CentralConfCondition}
\bm r_a - \bm r_\st{cm} \propto \frac{1}{m_a} \frac{\partial V_\st{New}}{ \partial \bm r_a} \,, \qquad   \bm r_\st{cm} = \frac{\sum_{a=1}^N m_a \bm r_a}{m_\st{tot}} \,,  \qquad m_\st{tot} = \sum_{a=1}^N m_a \,.
\end{equation}
\end{Theorem}

Another useful characterization of central configurations is as the stationary points of the \emph{complexity function}, also known as (minus) the shape potential or the normalized Newton potential:
\begin{equation}
C_\st{S} (\bm r_a) = - m_\st{tot}^{-3/2} \frac{V_\st{New}}{\sqrt{I_\st{cm}}} \,, \qquad I_\st{cm} =  \sum_{a=1}^N m_a \| \bm r_a - \bm r_\st{cm} \|^2 =  \sum_{a<b} m_a m_b \| \bm r_a - \bm r_b \|^2 \,.
\end{equation}

It is easy to check that Equation~(\ref{CentralConfCondition}) follows from $ \frac{\partial C_\st{S}}{ \partial \bm r_a}  =0 $.

We will sketch here Chazy's proof of the theorem. It uses three fundamental equalities, valid for any homogeneous $N$-body potential $U = U(\bm r_a)$:

\begin{enumerate}
\item \textbf{{Conservation of energy}.} The following quantity is a constant of motion:
\begin{equation}\label{EnergyConservation}
E = T + U \,,
\end{equation}
where $T = \sum_{a=1}^N \frac{\| \bm p_a \|^2}{2 m_a}$ is the total kinetic energy of the~system.\\

\item \textbf{{Lagrange--Jacobi relation}.} A first version of this equation has been given by Lagrange~\cite{Lagrange1772}. If~the potential  is homogeneous of degree $k$, i.e.,~$U(\alpha \bm r_a) = \alpha^k U(\bm r_a)$ for any real positive constant $\alpha$, then
\begin{equation}\label{LagrangeJacobiRelation}
\ddot I_\st{cm} = 4 (E -  U ) - 2 k U \,.
\end{equation}

This identity can be proved using Euler's homogeneous function theorem, which states that $\sum_{a=1}^N \bm r_a \cdot  \frac{\partial U}{ \partial \bm r_a} = k U$.
In the case of the Newtonian potential $k=-1$, and~this equality turns into
\begin{equation}\label{LagrangeJacobiRelationNewtonsPotential}
\ddot I_\st{cm} = 4 E  - 2 U \,.
\end{equation}

Notice that,  in~the case of Newton's potential, since $4 E  - 2 U $ and $U<0$, the~Lagrange--Jacobi relation implies that $\ddot I_\st{cm} >0$ if $E\ge 0$. So the moment of inertia is either a U-shaped function, going through a minimum and growing monotonically in the two time directions away from it, or~it has a zero at a certain instant $t=0$ and is defined only on one side of $t=0$, growing monotonically away from~it.

It follows directly from the Lagrange--Jacobi equation that, for~$E\ge 0$, a~total collision can only occur at the minimum of the $I$-curve, where $\dot{I}_{cm}=0$. Let us, at~this point, introduce the notion of the dilatational momentum $D = \sum_{i=1}^N \bm r_a \cdot \bm p_a$. We find that $D= 1/2 \dot{I}_{cm}$ and, thus, a~total collision can only occur at $D=0$.\\

\item \textbf{{Chazy's kinetic energy decomposition theorem.}} One can write
\begin{equation}\label{KineticEnergyDecomposition}
T = T_\st{cm} + \frac{1}{2 I_\st{cm}}  \left( \frac 1 4  \dot I_\st{cm}^2 + \| \bm{L} \|^2 + 2 \, T_\st{S} \right)  \,,
\end{equation}
where $T_\st{S}$, the~\emph{shape kinetic energy}, is a sum of squares and therefore positive. The~above relation was rediscovered much later by Saari~\cite{Saari1984} as a consequence of his velocity decomposition theorem (which states that the center-of-mass motion, dilatation, rotation, and~shape components of the velocity $3N$-vector are orthogonal). The~above relation is also at the basis of Sundman's inequality $2T \geq \frac{1}{I_\st{cm}}  \left( \frac 1 4  \dot I_\st{cm}^2 + \| \bm{L} \|^2 \right) $, proved in 1912~\cite{Sundman1912}. 

\end{enumerate}

Combining Equations~(\ref{EnergyConservation}) and~(\ref{LagrangeJacobiRelationNewtonsPotential}) we get $\ddot I_\st{cm} = 2 E +  2 T$, and~using~(\ref{KineticEnergyDecomposition}) we can remove the total kinetic energy and get
\begin{equation}
\ddot I_\st{cm} - \frac{\dot I_\st{cm}^2}{4 I_\st{cm}} = \frac{ \| \bm{L} \|^2}{ I_\st{cm}} + \frac{2 T_\st{S}}{I_\st{cm}}  +  2 (E+T_\st{cm}) \,.
\end{equation}

The left-hand side can be rewritten as $ \frac{I_\st{cm}^{1/2} }{\dot I_\st{cm}} \frac{d}{dt} \left( \frac{\dot I_\st{cm}^2}{2 I_\st{cm}^{1/2} } \right)$, and~then the equation takes the following form:
\begin{equation}
\frac{d}{dt} \left( \frac{\dot I_\st{cm}^2}{2 I_\st{cm}^{1/2} } \right) = \| \bm{L} \|^2 \frac{\dot I_\st{cm}}{ I_\st{cm}^{3/2}}  +  2 (E+T_\st{cm}) \frac{\dot I_\st{cm}}{I_\st{cm}^{1/2}} + \frac{2 T_\st{S} \dot I_\st{cm}}{I_\st{cm}^{3/2}} \,.
\end{equation}

Using the fact that $E$, $T_\st{cm}$, and $\bm L$ are conserved quantities, we can integrate the above equation in $d t$ from $t_0$ to $t$:
\begin{equation}
\frac{\dot I_\st{cm}^2}{2 I_\st{cm}^{1/2} }  = -\frac{2 \| \bm{L} \|^2 }{I_\st{cm}^{1/2}}+  4 (E+T_\st{cm})I_\st{cm}^{1/2}   +  \int_{t_0}^{t} \frac{2 T_\st{S} \dot I_\st{cm}}{I_\st{cm}^{3/2}} dt + \text{\it const} \,.
\end{equation}

Now, {{suppose} that $\dot I_\st{cm} \leq 0$ over the whole interval of  integration} (by what was said above, in~the case we are interested in, $I_\st{cm}$ goes to zero monotonically and is not defined past it), then, since $T_\st{S} >0$, $ \int_{t_0}^{t} \frac{2 T_\st{S} \dot I_\st{cm}}{I_\st{cm}^{3/2}} dt <0$. The~angular-momentum term $-\frac{2 \| \bm{L} \|^2 }{I_\st{cm}^{1/2}}$ is negative or zero, and~ because $\bm{L}$ is conserved, it either diverges to $-\infty$ as $I_\st{cm} \to 0$, or~it stays zero the whole time, if~the total angular momentum is zero. The~term $4 (E+T_\st{cm})I_\st{cm}^{1/2}$ vanishes as $I_\st{cm} \to 0$, and~the integration constant stays constant. Therefore, we have an equation with the structure:
\begin{equation}
\frac{\dot I_\st{cm}^2}{2 I_\st{cm}^{1/2} } - \int_{t_0}^{t} \frac{2 T_\st{S} \dot I_\st{cm}}{I_\st{cm}^{3/2}} dt  = -\frac{2 \| \bm{L} \|^2 }{I_\st{cm}^{1/2}} +  f(t) \,.
\end{equation}
where the left-hand side is positive-definite, and~$ f(t)$ tends to a finite constant as $I_\st{cm}  \to 0$. Therefore, the only way that this identity can be preserved all the way to an instant in which  $I_\st{cm}$ vanishes, is that {{the total angular momentum has to be zero.}} 

Then we are left with the sum of two positive quantities:
\begin{equation}
\frac{\dot I_\st{cm}^2}{2 I_\st{cm}^{1/2} } - \int_{t_0}^{t} \frac{2 T_\st{S} \dot I_\st{cm}}{I_\st{cm}^{3/2}} dt \,,
\end{equation}
which is equal to a function that remains finite when $I_\st{cm}  \to 0$. Each of these quantities then have to admit a finite limit at a total~collision.

Let us focus on the first of those two quantities. Its square root, $\frac{\dot I_\st{cm}}{ I_\st{cm}^{1/4} }$, will admit a finite limit too. Calling ${\displaystyle \lim_{t\to 0 }} \left( \frac{\dot I_\st{cm}}{2 I_\st{cm}^{1/4} } \right) = \ell $, where $t=0$ is the time of total collision, we can then~write
\begin{equation}\label{Idot_asymptotics}
\frac{\dot I_\st{cm}}{2 I_\st{cm}^{1/4} } = \ell + \epsilon(t) \,,
\end{equation}
where $\epsilon(t) \xrightarrow[t\to 0]{} 0$. Since $\frac{\dot I_\st{cm}}{I_\st{cm}^{1/4} } = \frac 4 3 \frac{d}{dt}  I_\st{cm}^{3/4}$, we can integrate the equation above as
\begin{equation}\label{I_asymptotics}
I_\st{cm}^{3/4} =  \left( \frac 3 2 \, \ell  + \delta(t) \right) t \,,
\end{equation}
where $\delta(t) \xrightarrow[t\to 0]{} 0$. We conclude that, {{if a central collision happens at} $t=0$, the~quantity:
\begin{equation}\label{DefJ}
J = \frac{I_\st{cm}}{t^{4/3}} \,,
\end{equation} {admits a finite limit as} $t \to 0$.}

Consider now the following transformation:
\begin{equation}
\bm r_a = \bm r_\st{cm} +  t^{2/3} \, \bm s_a \,.
\end{equation}

The new variables are subject to the following equations of motion:
\begin{equation}
t^{2/3} \ddot {\bm s}_a  + \frac 4 3 t^{-1/3} \dot {\bm s}_a - \frac 2 9 t^{-4/3} {\bm s}_a  = -\frac{t^{-4/3}}{m_a} \frac{\partial V_\st{New} ({\bm s}_b)}{\partial {\bm s}_a  }, 
\end{equation}
where $V_\st{New} ({\bm s}_b)$ is Newton's potential, with~$\bm r_a $ replaced with $\bm s_a$. These equations can be rewritten in an autonomous form, by~reparametrizing time with a logarithm, $u = - \log t$, which goes to $-\infty $ at the total collision $t =0$:
\begin{equation}\label{EqsOfMotion_s_variables}
 {\bm s}''_a   =  \frac 1 3  {\bm s}'_a + \frac 2 9  {\bm s}_a  - \frac{1}{m_a} \frac{\partial V_\st{New} ({\bm s}_b)}{\partial {\bm s}_a  } \,,
\end{equation}
where $f'(u) = \frac{d f(u)}{du}$. Consider now the quantity $J$ defined in~(\ref{DefJ}). We established already that in the $t \to 0$ ($u \to +\infty$) limit, $J$ tends to a constant (which by definition cannot be positive) in the time interval of interest). Now we shall prove that this constant cannot be zero. Consider first the  $u$-derivative of $J$. We can prove that it vanishes at the total collision. In~fact:
\begin{equation}
J' = -t \dot J = - t \left(  \frac{\dot I_\st{cm}}{t^{4/3}} - \frac 4 3 \frac{I_\st{cm}}{t^{7/3}} \right)
=  \frac 4 3 \frac{I_\st{cm}}{t^{4/3}} -\frac{\dot I_\st{cm}}{t^{1/3}}  \,.
\end{equation}
At this point we can use Equations~(\ref{Idot_asymptotics}) and~(\ref{I_asymptotics}), which imply that
\begin{equation}
\begin{aligned}
\dot I_\st{cm} &= 2  \left( \ell + \epsilon(t)\right) \left( \frac 3 2 \, \ell  + \delta(t)  \right)^{1/3} t^{1/3}  \,,
\\
I_\st{cm} &=  \left( \frac 3 2 \, \ell  + \delta(t) \right)^{4/3} t^{4/3}  \,,
\end{aligned}
\end{equation}
where $\epsilon$ and $\delta$ vanish at $t=0$. So
\begin{equation}
J' =   \frac 4 3 \left( \frac 3 2 \, \ell  + \delta(t) \right)^{4/3} - 2  \left( \ell + \epsilon(t)\right) \left( \frac 3 2 \, \ell  + \delta(t)  \right)^{1/3}  \xrightarrow[t \to 0]{}    0  \,.
\end{equation}

Now consider the kinetic energy $T = \frac 1 2 \sum_a m_a \|\dot{\bm r}_a \|^2$. We can express it in terms of the $\bm s_a$ variables and their $u$-derivatives  as
\begin{equation}
T = t^{-2/3}  \left[ \frac 2 9 J - \frac 1 3 J' + \frac 1 2 \sum_a m_a \left\|\frac{d \bm s_a}{du} \right\|^2 \right] \,,
\end{equation}
and, if~we call $S = \frac 1 2 \sum_a m_a \|\frac{d \bm s_a}{du} \|^2$, we can then rewrite the  energy conservation equation and the Lagrange--Jacobi relation in its two forms as
\begin{equation}\label{TheThreeEquations}
\begin{gathered}
\frac 2 3 J' - \frac 4 9 J = 2 S + 2 V_\st{New}(\bm s_a) - 2 E e^{-\frac 2 3 u} \,,
\\ 
J'' - \frac 5 3 J' + \frac 4 9 J = - 2  V_\st{New}(\bm s_a) + 4 E e^{-\frac 2 3 u} \,,
\\
J'' - J' =  2 S + 2 E e^{-\frac 2 3 u} \,.
\end{gathered}
\end{equation}

Note, at~this point, that if $ J \xrightarrow[u \to \infty]{} 0$, then the rescaled Newton potential would diverge,~because~\begin{equation}
J = \frac{I_\st{cm}}{t^{4/3} } = \sum_a m_a \| \bm r_a - \bm r_\st{cm} \|^2 t^{-4/3} = \sum_a m_a \| \bm s_a - \bm s_\st{cm} \|^2 \,,
\end{equation}
and, hence,
\begin{equation}J = 0  \Leftrightarrow \bm s_a = \bm s_b ~~ \forall a,b \,,
\end{equation}
which implies that $V_\st{New}(\bm s_a) = - \infty$. Therefore, if~ $ J \xrightarrow[u \to \infty]{} 0$,  the~second of the Equation~(\ref{TheThreeEquations}) tells us that $ J'' \xrightarrow[u \to \infty]{} +\infty $.  $J''$ going to the definite limit $+ \infty$ is impossible since the first derivative converges to zero, $ J' \xrightarrow[u \to \infty]{} 0$. ({{If} $ J'' \xrightarrow[u \to \infty]{} +\infty $, there exists $u_1 >0$, such that $J''(u)>J''(u_1) >0$ for all $u > u_1$. Then the integral $J'(u) - J'(u_1) = \int_{u_1}^u J''(u) du > J''(u_1) \int_{u_1}^u  du $, which tends to infinity as $u \to + \infty$, and~it is impossible for $J'$ to tend to a finite value at {infinity}}.)  Therefore $J$ cannot go to zero at the total~collision.

This proves that $J$ tends to a strictly positive finite value at the total collision.
Now consider the third of the Equation~(\ref{TheThreeEquations}). We can integrate it with respect to $u$ over an  interval beginning at $u$:
\begin{equation}
J'(u) - J(u) =  2 \int_{u}^{u_1} S \, du  - 3 E e^{- \frac 2 3 u} + \text{\it const.}
\end{equation}

Since $J'(u) \to 0$ as $u \to \infty$, and~$J(u)$ tends to a finite constant, the~integral $\int_{u_0}^{\infty} S \, du $ is finite. However,~the integrand is  equal to the sum of squares:
\begin{equation}
\int_{u_0}^{\infty} S \, du = \frac 1 2 \sum_a m_a \int_{u_0}^{\infty} \left\|\frac{d \bm s_a}{du} \right\|^2 \, du \,,
\end{equation}
and Chazy can now prove that $\frac{d s^i_a}{du}$ all go to zero at infinity using the fact that $\int_{u_0}^{\infty} S \, du $ is finite and that the logarithmic derivative
\begin{equation}\label{PreEqsOfMotion}
\frac{\frac{d^2 s^i_a}{du^2}}{\frac{d s^i_a}{du}} = \frac 1 3  + \frac{\frac 2 9 s^i_a - \frac{\partial V_\st{New} ({\bm s}_b)}{\partial s^i_a} }{\frac{d s^i_a}{du}}
\end{equation}
is~bounded.

If $\frac{d s^i_a}{du}$ all go to zero at $u \to \infty$, then the first of the Equation~(\ref{TheThreeEquations}) implies that $V_\st{New}(\bm s_a) $ attains a finite limit there. Therefore, the partial derivatives (of all orders) of $V_\st{New}(\bm s_a) $ with respect to $\bm s_a$ are bounded, and~so are, from~Equation~(\ref{EqsOfMotion_s_variables}), all accelerations $\frac{d^2 s^i_a}{d u^2}$. Differentiating Equation~(\ref{EqsOfMotion_s_variables}) with respect to $u$, we get
\begin{equation}
 {\bm s}'''_a   = \frac 1 3  {\bm s}''_a + \frac 2 9  {\bm s}'_a  - \frac{1}{m_a} \frac{\partial^2 V_\st{New} ({\bm s}_b)}{\partial {\bm s}_a \partial {s}^j_b } {s'}^j_b\,,
\end{equation}
which implies that ${\bm s}'''_a $ is bounded, too. Chazy now quotes a theorem by Hadamard stating that \emph{``when a function goes to a finite limit at infinity, and~its  second derivative is bounded, then its first derivative vanishes at infinity''}.

This theorem implies that $\frac{d^2 s^i_a}{du^2} \xrightarrow[u \to  \infty]{} 0$, and, since the first derivative vanishes as well, the~equations of motion~(\ref{EqsOfMotion_s_variables}) imply that, asymptotically,
\begin{equation}
\frac 2 9  {\bm s}_a  =  \frac{1}{m_a} \frac{\partial V_\st{New} ({\bm s}_b)}{\partial {\bm s}_a }\,.
\end{equation}

This is identical to the central configuration condition~(\ref{CentralConfCondition}), with~proportionality factor $\frac 9 2 t^{2/3}$.

\section{Total Collisions on Shape~Space}\label{sec3}

\subsection{Phase Space Reduction of the Planar Three-Body~Problem}
\label{3BodyPhaseSpaceReduction_sec}

Here we recount the elimination, from~the 3-body problem, of~the extrinsic degrees of freedom, i.e.,~those that have to do with the position and orientation of the system in absolute space. We will focus on the planar case, that is, when the total angular momentum is orthogonal to the plane of the three particles, which means that the particles never leave that plane and the treatment is simplified. The~zero angular momentum case, which we are ultimately interested  in, can be seen as a particular case of the planar one. Our treatment will follow the one of~\cite{Barbour:2015}.

The extended phase space of the Newtonian three-body problem is $\mathbbm R^{18}$, with~coordinates ${\bf r}_1, {\bf r}_2, {\bf r}_3 \in \mathbbm R^3$ for the particle positions, and~${\bf p}^1, {\bf p}^2, {\bf p}^3 \in \mathbbm R^3$ for the momenta.
It is well known that if the angular momentum ${\bf L} = \sum_{a=1}^3 {\bf r}_a \times {\bf p}^a$ is orthogonal to the plane identified by the three particles, ${\bf L} \times \left( ({\bf r}_1 -{\bf r}_3)\times ({\bf r}_2 -{\bf r}_3)\right) =0 $, then the three particles never leave that plane during the~evolution. 

Let us assume, from~now on, that the problem is planar.
We can then assume that the position and momenta are two-component vectors  ${\bf r}_1, {\bf r}_2, {\bf r}_3 ,{\bf p}^1, {\bf p}^2, {\bf p}^3 \in \mathbbm R^2$. The~degrees of freedom are six, but~three are gauge, corresponding to the translations and rotations on the plane of the motion. We have to restrict to the hypersurface ${\bf P}=L_\perp=0$, where $L_\perp$ is the remaining non-zero component of the angular momentum, and~then we have to quotient by the transformations generated by these constraints. It turns out that in this case we can take the `royal road' of explicitly identifying a sufficient number of gauge-invariant degrees of freedom (observables), and~perform a coordinate transformation in phase space that separates them from the gauge degrees of freedom, making them orthogonal~coordinates.

To deal with translations, we define the mass-weighted Jacobi coordinates:
\begin{equation}
\begin{aligned}
\bm \rho_1 &= \textstyle \sqrt{ \frac{m_1 \, m_2 }{m_1 + m_2} }\left( \mathbf  r_2 - \mathbf  r_1 \right), \\
 \bm \rho_2 &=\textstyle \sqrt{ \frac{m_3 \, (m_1+m_2) }{m_1 + m_2 + m_3} } \left({\bf r}_3 - \frac{m_1 \, {\bf r}_1 + m_2 \, {\bf r}_2}{m_1+m_2} \right),\\
 \bm \rho_3 &= \textstyle \frac{1}{\sqrt{m_1+m_2+m_3}} \left( m_1\, \mathbf  r_1 + m_2 \, \mathbf  r_2 + m_3 \, \mathbf  r_3 \right).
\end{aligned}
\end{equation}

The transformation to them is linear and invertible,
\begin{equation}
\bm \rho_a = {M_a}^b \, {\bf r}_b \,,\qquad  \det M = \sqrt{m_1 m_2 m_3},
\end{equation}
so, looking at the symplectic potential,
\begin{equation}
\Theta = \bm p^a \d \bm r_a = \bm p^a (M^{-1})_a{}^b \bm \rho_b = \bm \kappa^a \bm \rho_a \,,
\end{equation}
it appears obvious that the momenta conjugate to $\bm \rho_a$ are related to $\bm p^a$ by the transpose of the inverse of the matrix $M$ (notice that $M$ is not symmetric):
\begin{equation}
\bm \kappa^a = (M^{-1})_b{}^a \, \mathbf p^b = ((M^{-1})^\st{T})^a{}_b \, \mathbf p^b \, .
\end{equation}

The inverse transformation is (the transpose and the inverse of an invertible matrix~commute):
\begin{equation}
{\bf p}^a = {M^a}_b \, \bm \kappa^b = {(M^\st{T})_b}^a \,\bm \kappa^b.
\end{equation}

Note that the inverse matrix is
\begin{equation}
{(M^{-1})^a}_b =\textstyle \frac{1}{\sqrt{m_1+m_2+m_3}} \left(
\begin{array}{ccc}
 -\sqrt{\frac{m_2 \left(m_1+m_2+m_3\right)}{m_1 \left(m_1+m_2\right)}} &  \sqrt{\frac{m_1 \left(m_1+m_2+m_3\right)}{m_2 \left(m_1+m_2\right)}} & 0 \\
 -\sqrt{\frac{m_3}{m_1+m_2}}  & -\sqrt{\frac{m_3}{m_1+m_2}} & \sqrt{\frac{m_1+m_2}{m_3}} \\
 1 & 1 & 1 \\
\end{array}
\right)\,
\end{equation}
and has a constant column. It is the column of $\bm \kappa^3$,
\begin{equation}
\bm \kappa^3 = \frac{1}{\sqrt{m_1+m_2+m_3}} \sum_{a=1}^3 \mathbf p^a \,,
\end{equation}
which is, therefore, proportional to the total momentum and decouples from the problem. The~coordinates
$\bm \rho_3$ are the coordinates of the center of mass, which decouple too.
The other two momenta are
\begin{equation}
\bm \kappa^1 =  \frac{m_1 {\bf p}^2-m_2 {\bf p}^1}{\sqrt{m_1 m_2 \left(m_1+m_2\right)}}\,, \qquad \bm \kappa^2 =\frac{\left(m_1+m_2\right) {\bf p}^3-m_3 ({\bf p}^1+{\bf p}^2)}{\sqrt{\left(m_1+m_2\right) m_3 \left(m_1+m_2+m_3\right)}}.
\end{equation}

As we said, the~transformation to Jacobi coordinate and momenta is canonical, and,~therefore, it leaves the Poisson brackets invariant:
\begin{equation}
\{ \rho^i_a , \kappa^b_j \} = {\delta^i}_j \,{\delta_b}^a.
\end{equation}

The kinetic term is diagonal in the momenta $\bm \kappa^a$,
\begin{equation}
T =  \sum_{a=1}^3 \frac{\mathbf p^a \cdot \mathbf p^a  }{2 \, m_a} =  \sum_{a=1}^3 \sum_{b,c=1}^3  \frac{{M^a}_b \, {M^a}_c   }{2 \, m_a}  \bm \kappa^b \cdot \bm \kappa^c   = \frac 1 2  \sum_{a=1}^3  \bm \| \bm \kappa^a \|^2,
\end{equation}
as is the moment of inertia,
\begin{equation}
I_\st{cm} = \sum_{a=1}^3 m_a \, \|  {\bf r}_a - {\bf r}_\st{cm} \|^2 = \sum_{a=1}^3 \sum_{b,c=1}^3  {(M^{-1})_a}^b \, {(M^{-1})_a}^c   m_a \,  \bm \rho_b \cdot \bm \rho_c  = \sum_{a=1}^2 \| \bm \rho_a\|^2 \,,
\end{equation}
(notice how the sum is from $a=1$ to $2$, because~$I_\st{cm}$ does not depend on the coordinates of the center of mass $\bm \rho_3$. The~inertia tensor also takes a particularly simple form:
\begin{equation}
\mathbbm{I}_\st{cm} =\sum_{a=1}^3 m_a\left( \mathbbm{1} \, {\bf r}_a^\st{cm}\cdot{\bf r}_a^\st{cm} - {\bf r}_a^\st{cm}\otimes{\bf r}_a^\st{cm} \right) 
= \sum_{a=1}^2 \left(   \mathbbm{1} \bm \rho^a \cdot \bm \rho^a - \bm \rho^a \otimes \bm \rho^a \right).
\end{equation}

We are left with four coordinates $\bm \rho_1$, $\bm \rho_2$ and momenta $\bm \kappa^1$, $\bm \kappa^2$, and~a single angular momentum component (the one perpendicular to the plane of the triangle):
\begin{equation}
L_\perp = \frac{{\bf L} \cdot \left( ({\bf r}_1 -{\bf r}_3)\times ({\bm r}_2 -{\bm r}_3)\right)}{\left\| ({\bm r}_1 -{\bm r}_3)\times ({\bm r}_2 -{\bm r}_3)\right\|} = \sum_{a=1}^2 \, (\bm \rho_a \times \bm \kappa^a) \,,
\end{equation}
where with the vector product between two 2-dimensional vector we understand a scalar ${\bf a} \times {\bf b} = a_x b_y - a_y b_x$. The~coordinates
\begin{equation}\label{w-coordinates}
w_1 = \frac 1 2 \left(||\bm \rho_1||^2 - || \bm \rho_2 ||^2 \right), \qquad
w_2 = \bm \rho_1 \cdot \bm \rho_2\,,  \qquad   w_3 = \bm \rho_1 \times \bm \rho_2
\end{equation}
are invariant under the remaining rotational symmetry and, therefore, give a complete coordinate
system on the reduced configuration space. Notice that $w_3$ changes sign under a planar reflection (changing the sign of one of the coordinates, say $x$, of~both ${\bm \rho}_1$ and ${\bm \rho}_2$) while $w_1$ and  $w_2$ remain invariant, and,~therefore, the map $w_3 \to - w_3$ relates triangles conjugate under mirror transformations. This also has the consequence that the $w_3=0$ plane contains only collinear configurations (whose mirror image is identical to the original, modulo a planar rotation). This has nothing to do with 3D reflections (obtained by changing the sign of \emph{all} components of every Euclidean vector). In~fact triangles are invariant under such parity transformations, because~their parity conjugate is related to the original by a non-planar~rotation.

\begin{figure}
\begin{center}
\includegraphics[width=0.7\textwidth]{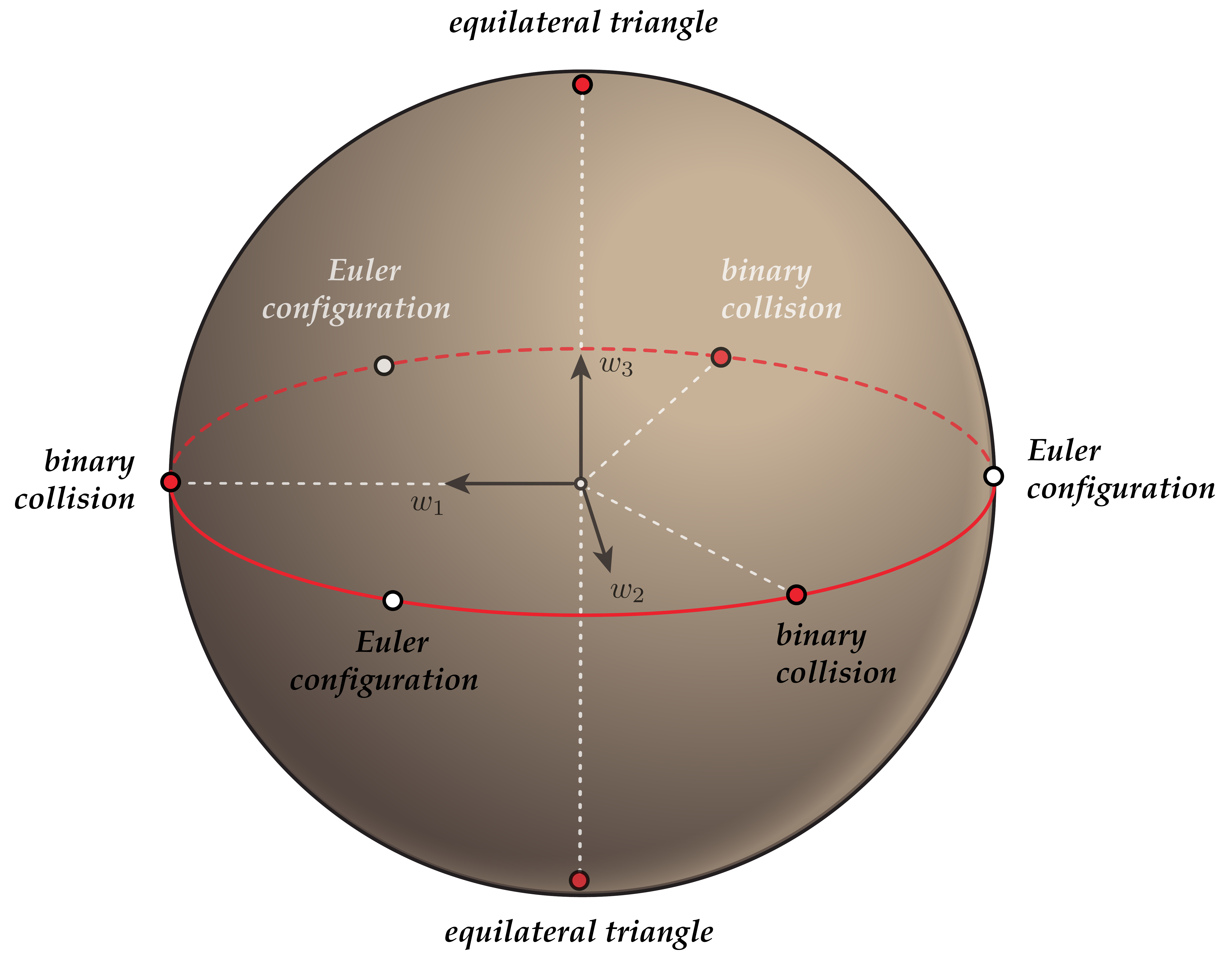}
\caption[The shape sphere]{\emph{The shape sphere of the equal-mass 3-body problem. Every point on the sphere
(defined as a constant-$\| \vec w \|$ surface) is a triangle. Points at the same longitude with opposite latitudes correspond to mirror-conjugated triangles. At~the poles (the intersections with the $w_3$ axis) we have the equilateral triangles, while on the equator (the red circle, $w_3=0$) we have the collinear configurations. Among~them, there are six special ones: three binary collisions (red dots, one of which is on the $w_1$ axis), and~Euler configurations (white dots), in~which the gravitational force acting on each particle points towards the center of mass and has a magnitude such that, if~the system is prepared in rest at one of these configurations, it will fall homothetically (without changing its shape) to a total collision at the centre of mass. The~same thing happens at the equilateral triangle (for all values of the masses, as~Lagrange showed).
Notice that the  Euler configurations and binary collisions are on the equator for all values of the three masses, but~their relative positions on the equator depends on the masses. The~equilateral triangles are at the poles only in the equal-mass case.}}
\label{ShapeSphereFigure}
\end{center}
\end{figure}

The Euclidean norm of the 3D vector $\vec w  = (w_1,w_2,w_3)$ is proportional  to (one quarter) the square of the moment of inertia \begin{equation}
||\vec w||^2  = \frac 1 4\left(||\bm \rho_1||^2 + || \bm \rho_2 ||^2 \right)^2 = \frac{I^2_\st{cm}} 4,
\end{equation}
so the angular coordinates in the three-space $(w_1,w_2,w_3)$ coordinatize shape space, which has the topology of a sphere~\cite{Mont}.  We call it the \emph{shape sphere}, and~in {Figure}~\ref{ShapeSphereFigure} we describe its salient~features.

The norms of the original Jacobi coordinate vectors can be written as
\begin{equation}
\| {\bm \rho}_1 \|^2 = \sqrt{w_1^2+w_2^2+w_3^2}+w_1
\,, \qquad \| {\bm \rho}_2 \|^2 = \sqrt{w_1^2+w_2^2+w_3^2}-w_1,
\end{equation}
and, therefore, the full vectors are specified by
\begin{equation}\label{JacobiCoordinatesVsWcoordinates}
\begin{aligned}
{\bm \rho}_1= \sqrt{ \sqrt{w_1^2+w_2^2+w_3^2}+w_1 } \left( \cos(\theta - \delta/2),\sin(\theta - \delta/2)\right)
,\\
{\bm \rho}_2= \sqrt{ \sqrt{w_1^2+w_2^2+w_3^2}-w_1 } \left( \cos(\theta + \delta/2),\sin(\theta + \delta/2)\right),
\end{aligned}
\end{equation}
where~$\theta = {\sfrac 1 2} \left[  \arctan  \left( \rho_1^y / \rho_1^x \right) + \arctan  \left( \rho_2^y  / \rho_2^x \right)\right]$ is an overall orientation angle which is not rotation-invariant and, therefore, is not fixed by the specification of the coordinates $w_1,w_2,w_3$; and $\delta = \arctan \frac{w_3}{w_2}$ is the angle between ${\bm \rho}_1$ and ${\bm \rho}_2$.
%

We now want to find the momenta conjugate to $\vec w$. To~do so, we consider the symplectic~potential
\begin{equation}
\Theta =  \sum_{a=1}^3 {\bf p}^a \cdot \d {\bm r}_a =
 \sum_{a=1}^3 {\bm \kappa}^a \cdot \d {\bm \rho}_a.
\end{equation}

If we replace ${\bm \rho}_a$ with their expressions in terms of $\vec w$ from Equation~(\ref{JacobiCoordinatesVsWcoordinates}), we get
\begin{equation}
\Theta = z^1 \, \d w_1 +z^2 \, \d w_2 +z^3 \, \d w_3 + {\bm \kappa}^3 \cdot \d {\bm \rho}_3  + L_\perp \, \d \theta,
\end{equation}
where
\begin{equation}
\begin{aligned}
&z^1= \textstyle   \frac{ {\bm \kappa}^1 \cdot {\bm \rho}_1 - {\bm \kappa}^2 \cdot {\bm \rho}_2 } {  \| {\bm \rho}_1\|^2 + \| {\bm \rho}_2\|^2},  \\
&z^2 = \textstyle \frac{  {\bm \kappa}^1 \cdot {\bm \rho}_2 + {\bm \kappa}^2 \cdot {\bm \rho}_1  }{  \| {\bm \rho}_1\|^2 + \| {\bm \rho}_2\|^2} - \frac 1 2  \frac{{\bm \rho}_1 \times {\bm \rho}_2}{\|{\bm \rho}_1\| \| {\bm \rho}_2\| } \frac {   \| {\bm \rho}_1\|^2 - \| {\bm \rho}_2\|^2 }{  \| {\bm \rho}_1\|^2 + \| {\bm \rho}_2\|^2} \left( {\bm \kappa}^1 \times {\bm \rho}_1 +   {\bm \kappa}^2 \times {\bm \rho}_2\right),  \\
&z^3 = \textstyle\frac{  \| {\bm \rho}_1\|^2  {\bm \kappa}^1 \times {\bm \rho}_2  -   \| {\bm \rho}_2\|^2 {\bm \kappa}^2\times {\bm \rho}_1 }{2  \| {\bm \rho}_1\|^2 \| {\bm \rho}_2\|^2 } - \frac 1 2  \frac{{\bm \rho}_1 \times {\bm \rho}_2}{\| {\bm \rho}_1\| \| {\bm \rho}_2\| } \frac {   \| {\bm \rho}_1\|^2 - \| {\bm \rho}_2\|^2 }{  \| {\bm \rho}_1\|^2 + \| {\bm \rho}_2\|^2}\left(  {\bm \kappa}^1 \cdot {\bm \rho}_1 - {\bm \kappa}^2 \cdot {\bm \rho}_2 \right), 
\\
& \theta = {\sfrac 1 2} \left[  \arctan  \left( \rho_1^y / \rho_1^x \right) + \arctan  \left( \rho_2^y  / \rho_2^x \right)\right],
\end{aligned}
\end{equation}
%
%
%
%
so we now have a complete canonical transformation from the coordinates (${\bm r}_1$, ${\bm r}_2$, ${\bm r}_3$; ${\bf p}^1$, ${\bf p}^3$, ${\bf p}^3$) to ($w_1$, $w_2$, $w_3$, ${\bm \rho}_3$, $\theta$; $z^1$, $z^2$, $z^3$, ${\bm \kappa}^3$, $L_\perp$). The~Poisson brackets in these coordinates are canonical,  as~they should be:
\begin{equation}
\begin{aligned}
\{ z^a , w_b \} = \delta^a{}_b,  \qquad \{ L_\perp , \theta\} = 1,\qquad \{ z^a , L_\perp \} = 0,
 \qquad \{ z^a, z^b\} =0,
\\
\{ z^a ,\kappa^3_j\} = 0,
 \qquad \{ \theta, \kappa^3_j\} =0, 
 \qquad \{ z^a ,\rho_3^j \} = 0,
 \qquad \{ \theta, \rho_3^j\} =0.
\end{aligned}
\end{equation}

In the new coordinates, the~kinetic energy decomposes as
\begin{equation}
T = \frac 1 2 \sum_{a=1}^3 \| {\bm \kappa}_a \|^2  =   \| \vec w  \| \left(  \| \vec z \|^2 + \frac{L_\perp^2}{4(w_2^2 + w_3^2)}\right) +  w_1 \left( \frac{ w_2 z^3 - w_3 z^2} {w_2^2 + w_3^2}\right) L_\perp + \frac 1 2 \| {\bm \kappa}^3 \|^2,
\end{equation}
and Newton's potential takes the form
\begin{equation}
V_\st{New} = - \sum_{a<b} \frac{ (m_a \, m_b)^{\frac 3 2}(m_a + m_b)^{-\frac 1 2}}{\sqrt{ \| \vec w  \|  -  w_1  \, \cos \, \phi_{ab} -  w_2  \, \sin \, \phi_{ab}}},
\end{equation}
where $\phi_{ab}$ are the longitudes on the shape sphere of the two-body collisions between particle $a$ and $b$. These can be found
by using Equation~(\ref{w-coordinates}):
\begin{itemize}
\item If $\bm r_1 = \bm r_2$, then $\bm \rho_1 \propto \bm r_2 - \bm r_1  =0$, and,~therefore, $w_2 =  \bm \rho_1 \cdot \bm \rho_2 = 0$, and~ $w_3 = \bm \rho_1 \times \bm \rho_2 =0$: we are on the equator and on the axis $1$. Moreover  $w_1 = \frac 1 2 \left(||\bm \rho_1||^2 - || \bm \rho_2 ||^2 \right) =  - \frac 1 2  || \bm \rho_2 ||^2  < 0$, so
\begin{equation}
\phi_{12} = \pi 
\,.
\end{equation}

\item If $\bm r_2 = \bm r_3$, then $ \bm \rho_2 =\sqrt{ \frac{m_3 \, (m_1+m_2) }{m_1 + m_2 + m_3} }  \frac{m_1}{m_1+m_2} \left( {\bm r}_2 - {\bm r}_1 \right) \propto \bm \rho_1$, and,~therefore, $w_3 \propto = \bm \rho_1 \times \bm \rho_2 = 0$ and we are on the equator.
The $1$ and $2$ coordinates have values $w_1 = \frac 1 2 \left(  \frac{m_1 \, m_2 }{m_1 + m_2} -  \frac{m_3 m_1^2}{(m_1 + m_2 + m_3)(m_1+m_2)} \right) \left\| {\bm r}_2 - {\bm r}_1 \right\|^2 $ and $w_2 =
\sqrt{ \frac{m_1 \, m_2  \, m_3}{m_1 + m_2 + m_3} }  \frac{m_1}{m_1+m_2} \left\| {\bm r}_2 - {\bm r}_1 \right\|^2$, and~so the corresponding longitude is:
\begin{equation}
\begin{aligned}
\phi_{23} &= \arctan \left( \frac{2\sqrt{ \frac{m_1 \, m_2  \, m_3}{m_1 + m_2 + m_3} }  \frac{m_1}{m_1+m_2} } {  \frac{m_1 \, m_2 }{m_1 + m_2} -  \frac{m_3 m_1^2}{(m_1 + m_2 + m_3)(m_1+m_2)} } \right)
\\&=
\arctan \left( \frac{2\sqrt{ m_1 \, m_2  \, m_3(m_1 + m_2 + m_3) }   } {   (m_1 + m_2 + m_3) m_2 - m_3 m_1 } \right)  \,.
\end{aligned}
\end{equation}

{\item  If $\bm r_1 = \bm r_3$, the~same reasoning above applies to show that we are on the equator, because~$ \bm \rho_2 = \sqrt{ \frac{m_3 \, (m_1+m_2) }{m_1 + m_2 + m_3} } \frac{m_2}{m_1+m_2}  \left( \mathbf  r_1 - \mathbf  r_2 \right)$ is parallel to $\bm \rho_1$. Then $w_1 = \frac 1 2$ \linebreak$ \left( \frac{m_1 \, m_2 }{m_1 + m_2}   -  \frac{m_3 m_2^2 }{(m_1 + m_2 + m_3)(m_1+m_2)} \right)\left\| {\bm r}_2 - {\bm r}_1 \right\|^2$, and~$w_2 =- \left( \sqrt{ \frac{m_1 \, m_2 \, m_3  }{m_1 + m_2 + m_3} } \frac{m_2}{m_1+m_2}   \right)\left\| {\bm r}_2 - {\bm r}_1 \right\|^2$, and~the longitude is then}
\begin{equation}
\begin{aligned}
\phi_{13} &= - \arctan \left( \frac{2 \sqrt{ \frac{m_1 \, m_2 \, m_3  }{m_1 + m_2 + m_3} } \frac{m_2}{m_1+m_2}}{ \frac{m_1 \, m_2 }{m_1 + m_2}   -  \frac{m_3 m_2^2 }{(m_1 + m_2 + m_3)(m_1+m_2)} }
 \right)
\\&=
- \arctan \left( \frac{2\sqrt{ m_1 \, m_2  \, m_3(m_1 + m_2 + m_3) }   } {   (m_1 + m_2 + m_3) m_1 - m_3 m_2 } \right)  \,.
\end{aligned}
\end{equation}

\end{itemize}

If we call $\phi$ the azimuthal and $\psi$ the polar angle on the shape sphere, then
\begin{equation}
\frac{w_1}{\| \vec w  \|} = \cos \phi \, \cos \psi \,, ~~
\frac{w_2}{\| \vec w  \|} = \sin \phi  \, \cos \psi \,, ~~
\frac{w_3}{\| \vec w  \|} = \sin \psi \,,
\end{equation}
and on the constraint surface ${\bm \kappa}^3 =0$ the Hamiltonian takes the form
\begin{equation}
H = \| \vec w  \| \left(  \| \vec z \|^2 + \frac{L_\perp^2}{4(w_2^2 + w_3^2)}\right) +  w_1 \left( \frac{ w_2 z^3 - w_3 z^2} {w_2^2 + w_3^2}\right) L_\perp   - \frac {C_\st{S}(\psi,\phi)}{ \sqrt{\| \vec w  \|}}   \,,
\end{equation}
where
\begin{equation}\label{Complexity}
C_\st{S}(\psi,\phi) = \sum_{a<b} \frac{ (m_a \, m_b)^{\frac 3 2}(m_a + m_b)^{-\frac 1 2}}{\sqrt{ 1  - \cos \psi  \cos (\phi - \phi_{ab}) }}
\end{equation}
is the 3-body ``complexity function'' (according to the nomenclature used in~\cite{Barbour:2013,Barbour:2014,Barbour:2015,Barbour:2016}).
Finally, a short calculation reveals that the dilatational momentum takes basically the same form in the new coordinates:
\begin{equation}
D = \sum_{a=1}^3 {\bm r}_a \cdot {\bf p}^a  = 2 \, \vec w \cdot \vec z + {\bm \kappa}^3 \cdot {\bm \rho}_3 \,.
\end{equation}

We now want to separate the scale and shape degrees of freedom. Let us use $r = \sqrt{\| \vec w \|}$ as our scale, and~the angles $\psi$ and $\phi$ as our shape coordinates. The~symplectic potential now takes the form
\begin{equation}
\Theta = p_r \, \d r + p_\phi \, \d \phi  + p_\psi \, \d \psi + L_\perp \d \theta \,,
\end{equation}
where
\begin{equation}
\begin{aligned}
p_r &= 2 r (z^1 \cos \psi \cos \phi+z^2 \cos \psi \sin \phi+z^3 \sin \psi) \,,
\\
p_\phi &= r^2 \cos \psi (z^2 \cos \phi-z^1 \sin \phi) \,,
\\
p_\psi &= r^2 (z^3 \cos \psi-\sin \psi (z^1 \cos \phi+z^2 \sin \phi)) \,,
\end{aligned}
\end{equation}
which can be inverted as
\begin{equation}
\begin{aligned}
z^1 &=   \frac{\cos \phi (p_r  r \cos \psi-2 p_\psi  \sin \psi)-2 p_\phi \sec \psi \sin \phi}{2 r^2} \,, \\
z^2 &=  \frac{\sin \phi (p_r  r \cos \psi-2 p_\psi  \sin \psi)+2 p_\phi \sec \psi \cos \phi}{2 r^2} \,,
\\
z^3 &=  \frac{p_r  r \sin \psi+2 p_\psi  \cos \psi}{2 r^2} \,,
\end{aligned}
\end{equation}
which puts the Hamiltonian  in the following form:
\begin{equation}
H = \frac 1 4   p_r^2 + \frac{K}{r^2} - \frac 1 r C_\st{S}(\psi,\phi) \,,
\end{equation}
where the kinetic term $K$ is a quadratic form in the momenta $p_\psi$, $p_\phi$ and $L_\perp$, which is positive definite for any value of $\psi$ and $\phi$:\begin{equation}\label{ShapeKineticEnergyWithL}\begin{aligned}
K = \frac 1 4  L_\perp \left( \frac{L_\perp-4 p_{\phi } \sin \psi \cos ^2 \phi -p_{\psi } (\cos \psi +\cos (3 \psi)) \sin (2 \phi )}{\sin ^2 \psi +\cos ^2 \psi  \sin ^2 \phi} \right) \\
 + p_\psi^2 + \cos^{-2} \psi \, p_\phi^2 \,.
\end{aligned}\end{equation}



In the zero angular momentum case $L_\perp =0$, the~equations of motion are:
\begin{equation}
\begin{aligned}
&\dot r = \frac 1 2 p_r \,&&   \dot p_r =  2 \frac{p_\psi^2 + \cos^{-2} \psi \, p_\phi^2 }{r^3} - \frac 1 {r^2} C_\st{S}(\psi,\phi) \,, 
\\
&\dot \phi = \frac{2}{r^2} \cos^{-2} \psi \, p_\phi \, && \dot p_\phi = \frac 1 r \partial_\phi C_\st{S}(\psi,\phi) \,,
\\
&\dot \psi = \frac{2}{r^2}  p_\psi \, && \dot p_\psi = - \frac{4}{r^2} \cos^{-3} \psi \sin \psi \, p_\phi^2 +   \frac 1 r \partial_\psi C_\st{S}(\psi,\phi) \,,
\end{aligned}
\end{equation}
and in Lagrangian form:
\begin{equation}
\begin{aligned}
&\ddot r =   \frac r 4 \left( \dot \psi^2 + \cos^2 \psi \, \dot \phi^2\right) - \frac 1 {2 r^2} C_\st{S}(\psi,\phi) \,, 
\\
&\ddot \phi = - \frac{2}{r} \dot \phi \, \dot r  -2 \sin \psi \, \dot \psi \dot \phi + \frac{2}{r^3} \cos^{-2} \psi \,  \partial_\phi C_\st{S}(\psi,\phi) \,,
\\
&\dot \psi = - \frac 2 r \dot \psi \dot r +  \frac{2}{r^2} \left(  - r^2 \cos \psi \sin \psi \, \dot \phi^2 +   \frac 1 r \partial_\psi C_\st{S}(\psi,\phi) \right) \,.
\end{aligned}
\end{equation}

Finally, in~the new coordinates the dilatational momentum is of the following form:
\begin{equation}
D = 2 \, \vec w \cdot \vec z = r \cdot p_r \,.
\end{equation}

\subsection{Total Collisions in the Zero-Energy 3-Body~Problem}

In the previous subsection we described the phase space reduction of the 3-body problem (in the planar case, i.e.,~$\bm L$ orthogonal to the plane of the three bodies)  to shape degrees of freedom plus scale (the square root of the moment of inertia).  We ended up with a spherical shape space coordinatized by  two angles, $\psi \in (-\pi/2,\pi/2)$ and $\phi = (0,2\pi)$, plus a global orientation angle $\theta = (0,2 \pi)$ and the scale $r= \sqrt{\frac 1 2 I_{cm}}$, as~well as their four conjugate momenta $p_\phi$, $p_\psi$,  $p_r$, and $L_\perp$ with canonical symplectic structure. If~we specialize to the zero angular momentum case $L_\perp =0$ (the only case we are interested in if we want to study total collisions), the~coordinate $\theta$ drops out of the problem too, because~it is cyclic, and~we are left with two shape-space coordinates plus one scale, and~their conjugate momenta.
Conservation of energy is expressed by the following constraint equation:
\begin{equation}\label{ZeroEnergyHamiltonianConstraint}
\frac 1 4   p_r^2 + \frac{K}{r^2}- \frac 1 r C_\st{S}(\psi,\phi) - E  = 0\,,
\end{equation}
where $E$ is a constant (the total energy of the system), $K$ is the shape kinetic energy, written in Equation~(\ref{ShapeKineticEnergyWithL}), and~$C_\st{S}(\psi,\phi)$ is what we have been calling (see~\cite{Barbour:2013,Barbour:2014,Barbour:2015,Barbour:2016}) the complexity function, as~defined in Equation~(\ref{Complexity}). $C_\st{S}(\psi,\phi)$ is positive-definite,  and~we will study the vicinity of one of its stationary points, of~coordinates $(\phi_0,\psi_0)$.
The equations of motion in Newtonian time for the scale degree of freedom are:
\begin{equation}\label{ScaleNewtonianEqs}
\dot r = \frac 1 2 p_r \,, \qquad   \dot p_r =  2 \frac{K}{r^3} - \frac 1 {r^2} C_\st{S}(\psi,\phi) \,,
\end{equation}
which proves that the dilatational momentum $D = r \, p_r$ is monotonic. Since $D$ is monotonic, we can use it as an internal time parameter $\tau$ with $\tau=D$.

Note that, already at this point, the Hamiltonian constraint, together with the fact that $\tau \to 0$ at total collisions (which follows from the Lagrange--Jacobi equation, see above), implies that total collisions can happen only if the angular momentum and the shape momenta are all zero. In~fact, multiplying~(\ref{ZeroEnergyHamiltonianConstraint}) by $r^2$, we obtain:
\begin{equation}
 \frac 1 4   \tau^2 + K-  r C_\st{S}(\psi,\phi) - E \, r^2 = 0  \,,
\end{equation}
which, in~the limit $r \to 0$ and $\tau \to 0$,  implies $K \to 0$, and~$K$ is a positive-definite quadratic form in $p_\psi$, $p_\phi$, and $L_\perp$~(\ref{ShapeKineticEnergyWithL}).

As mentioned above, given that $\tau$ is monotonic, we can use it as an internal time parameter. Now the evolution with respect to $\tau$, in~the zero-energy case $E=0$, is described by the shape Hamiltonian $H_S$, the~canonical conjugate of $\tau = D$ expressed in terms of the shape variables by means of the Hamiltonian constraint $H=0$ (we obtain a reduced Hamiltonian dynamics on shape space precisely because $D$, our new internal time parameter, is the dilatational momentum, i.e.,~the generator of scalings, cf.~\cite{Barbour:2013}). To~determine $H_S$, we thus demand $\{H_S, D\}|_{H=0}\stackrel{!}{=}1$, so that $H_S$ is the logarithm of the solution of $H=0$ with respect to $r$. With~$p_r$ replaced by $\tau /r$, it is:
\begin{equation}
H_S = \log \left( \frac 1 4   \tau^2 + p_\psi^2 + \cos^{-2} \psi \, p_\phi^2 \right) - \log C_\st{S}(\psi,\phi) \,.
\end{equation}

It follows that the equations of motion of the ``decoupled system'' are
\begin{equation}\label{DecoupledSystem}
\begin{aligned}
& \frac{d \phi}{d\tau} = \frac{ 2 \cos^{-2} \psi \, p_\phi }{\frac 1 4   \tau^2 + p_\psi^2 + \cos^{-2} \psi \, p_\phi^2} 
\,, && 
\frac{d p_\phi}{d\tau} =  \frac{\partial_\phi C_\st{S}(\psi,\phi)}{C_\st{S}(\psi,\phi)} \,,
\\
&\frac{d \psi}{d\tau} = \frac{2  p_\psi}{\frac 1 4   \tau^2 + p_\psi^2 + \cos^{-2} \psi \, p_\phi^2} \,, && 
\frac{d p_\psi}{d\tau} = -\frac{ 2 \cos^{-3} \psi \sin \psi \, p_\phi^2}{\frac 1 4   \tau^2 + p_\psi^2 + \cos^{-2} \psi \, p_\phi^2}    + \frac{\partial_\psi C_\st{S}(\psi,\phi)}{C_\st{S}(\psi,\phi)} \,.
\end{aligned}
\end{equation}

We would now like to impose that the system undergoes a total collision. By~what we have seen in the previous section, this can only happen at a central configuration (the stationary points of $C_\st{S}(\psi,\phi)$), and~with vanishing dilatational momentum ($\tau=D=0$). However, imposing that the solution goes through a central configuration at the instant $\tau =0$ is not enough: it could simply be reaching a minimum of the moment of inertia (a Janus point) with the shape of a central configuration, and, past this minimum, grow again without ever hitting a total collision. In~order to get an actual total collision, the~moment of inertia has to vanish, that is, we need to have that $r \to 0$. However, according to \mbox{Equation~(\ref{DecoupledSystem})}, $r$ is no longer part of our description of the system (at this level of description, we already are on shape space). The~problem is now: if all I have is \mbox{system~(\ref{DecoupledSystem})}, how can I tell whether I reached a total collision or simply a $r\neq 0$ Janus point with the shape of a central configuration?  Is there a `manifest cause' for a total collision, which can be read off the curve on shape space?

The answer is yes: if some shape momenta $p_\psi$ and $p_\phi$ are non-zero at a central configuration, Equation~(\ref{DecoupledSystem}) tend to those of a spherical geodesic (in case the central configuration is on the equator of our coordinate system, the~term associated to the non-zero Christoffel symbols of the spherical metric on shape space, $\frac{ 2 \cos^{-3} \psi \sin \psi \, p_\phi^2}{\frac 1 4   \tau^2 + p_\psi^2 + \cos^{-2} \psi \, p_\phi^2}  $, vanishes, and~the equations reduce to those of a straight line). Otherwise, if~ both $p_\psi =0$ and $p_\phi =0$,  \mbox{Equation~(\ref{DecoupledSystem})} appear to diverge.  Indeed, it turns out that at a total collision the shape momenta must vanish (compare the remark above). Reconsider the Hamiltonian \mbox{constraint~(\ref{ZeroEnergyHamiltonianConstraint})} and multiply it by $r^2$:
\begin{equation}
p_\psi^2 + \cos^{-2} \psi \, p_\phi^2  = r^2 \, E -  \frac 1 4  (r p_r)^2 + r \,C_\st{S}(\psi,\phi)\,.
\end{equation}
We know, from~the discussion of the previous sections, that the dilatational momentum vanishes at a total collision, and,~therefore, $\tau = r p_r \to 0$. Moreover, the~complexity function remains bounded, and~the quantity $E$ is a constant of motion, so, in~the limit $r \to 0$, $p_\psi^2 + \cos^{-2} \psi \, p_\phi^2 $ must vanish, which implies $p_\psi =0$ and $p_\phi =0$ (cf. Reichert~\cite{Reichert2021}). This is a remarkable result in itself, for~it tells us that there exists a unique description of total collisions on (scale-free) shape~space.

We conclude that, in~order to discuss a total collision, we need to focus on those solutions of Equation~(\ref{DecoupledSystem}) which are perfectly tuned to reach a central configuration with exactly zero shape momenta. Let us now expand Equation~(\ref{DecoupledSystem}) in the vicinity of a central configuration $\phi = \phi_0 + \delta \phi$, $\psi = \psi_0 + \delta \psi$ , and~assume for simplicity that our coordinate system places this central configuration on the equator ({i.e.,} $\psi_0 =0$):
\begin{equation}\label{DecoupledSystemApproximated}
\begin{aligned}
& \frac{d \delta \phi}{d\tau} \sim \frac{ 2  \, p_\phi }{\frac 1 4   \tau^2 + p_\psi^2 + p_\phi^2} 
\,, && 
\frac{d p_\phi}{d\tau} \sim   H_{\phi,\phi} \delta \phi 
+ H_{\phi,\psi} \delta \psi \,,
\\
&\frac{d \delta \psi}{d\tau} \sim  \frac{2  p_\psi}{\frac 1 4   \tau^2 + p_\psi^2 + p_\phi^2} \,, && 
\frac{d p_\psi}{d\tau} \sim  -\frac{ 2  p_\phi^2}{\frac 1 4   \tau^2 + p_\psi^2 +  p_\phi^2} \delta \psi   + H_{\psi,\phi} \delta \phi + H_{\psi,\psi} \delta \psi \,,
\end{aligned}
\end{equation}
where $H_{ij} =  \left. \partial_i \partial_j \log C_\st{S} \right|_{\psi = \psi_0, \phi = \phi_0}$ are the components of the Hessian matrix of the logarithm of the complexity function at the central~configuration.

Now we can assume that $p_\phi$ and $p_\psi$ are small, too, as~we want to focus on a total collision which will make them vanish at $\tau =0$, so we can write $p_\phi = 0 + \delta p_\phi$ and $p_\psi = 0 + \delta p_\psi$ and expand at first order in $\delta p_i$:
\begin{equation}\label{DecoupledSystemApproximated2}
\begin{aligned}
& \frac{d \delta \phi}{d\tau} \sim \frac{ 8  \, \delta p_\phi }{ \tau^2 } 
\,, \qquad && 
\frac{d  \delta p_\phi}{d\tau} \sim   H_{\phi,\phi} \delta \phi + H_{\phi,\psi} \delta \psi \,,
\\
&\frac{d \delta \psi}{d\tau} \sim  \frac{8 \delta  p_\psi}{\tau^2 } \,, \qquad  && 
\frac{d  \delta p_\psi}{d\tau} \sim     H_{\psi,\phi} \delta \phi + H_{\psi,\psi} \delta \psi \,.
\end{aligned}
\end{equation}

This last step killed the Christoffel term, and~gave us a set of linear equations that can be diagonalized and~solved.
 
 \subsection{Asymptotics of Total-Collision~Solutions}

Equation~(\ref{DecoupledSystemApproximated2}) can be diagonalized. Let $\lambda_i$ be the $i$-th eigenvalue of the Hessian matrix $H$ with components $H_{ij} =  \left. \partial_i \partial_j \log C_\st{S} \right|_{\psi = \psi_0, \phi = \phi_0}$. Then, $H=T^{-1}\Lambda T$ where $\Lambda$ is the diagonalized matrix (with eigenvalues $\lambda_i$ as diagonal entries) and $T$ is composed of the normalized eigenvectors. Multiply Equation~(\ref{DecoupledSystemApproximated2}) from the left with $T$ and you obtain:
\begin{equation}
\frac{d  \rho_i}{d\tau} = \frac{8}{ \tau^2 } \pi_i
\,, \qquad
\frac{d   \pi_i}{d\tau} =  \lambda_i \rho_i \,,
\end{equation}
with
\begin{equation}
\rho_i = T_i{}^j\left(\begin{matrix}\delta \phi\\ \delta \psi \end{matrix}\right)_j \,,
\qquad
\pi_i = T_i{}^j \left(\begin{matrix}\delta p_\phi\\ \delta p_\psi \end{matrix}\right)_j \,.
\end{equation}

As a system of first-order ODEs, the~above clearly does not satisfy the Picard--Lindel\"of theorem at $\tau =0$: the right-hand side of the first equation is not continuous there, let alone~Lipshitz-continuous.

To solve the above equations, multiply the first by  $\tau^2$ and differentiate:
\begin{equation}
\tau^2 \frac{d^2  \rho_i}{d\tau^2}  + 2 \tau \frac{d  \rho_i}{d\tau} = 8 \, \frac{d  \pi_i}{d\tau} \,.
\end{equation}
and, replacing the second equation to eliminate $\pi_i$:
\begin{equation}
\tau^2 \frac{d^2  \rho_i}{d\tau^2}  + 2 \tau \frac{d  \rho_i}{d\tau} -  8\lambda_i \rho_i = 0 \,.
\end{equation}

Now, the~solutions of an equation of this form can be looked in the form of a monomial $A \, \tau^c$, which, when replaced in the equation, leads to the characteristic polynomial equation:
\begin{equation}
c (c-1)  + 2 \, c   - 8 \,  \lambda_i  = 0 \,.
\end{equation}

The above equation admits two solutions: $c^\pm(\lambda_i) = -\frac{1}{2} \pm \sqrt{\frac{1}{4} 
+ 8 \, \lambda_i}$, and,~therefore, the general solution of the differential equation is:
\begin{equation} \label{RhoSolution}
\rho_i = A^+_i  \, \tau^{c^+(\lambda_i)} + A^-_i  \, \tau^{c^-(\lambda_i)} \,.
\end{equation} 

We plot here the real part of $c^\pm$ vs. $\lambda$ (Figure~\ref{Figure_C}):
\begin{figure}
\begin{center}
\includegraphics[width=0.7\textwidth]{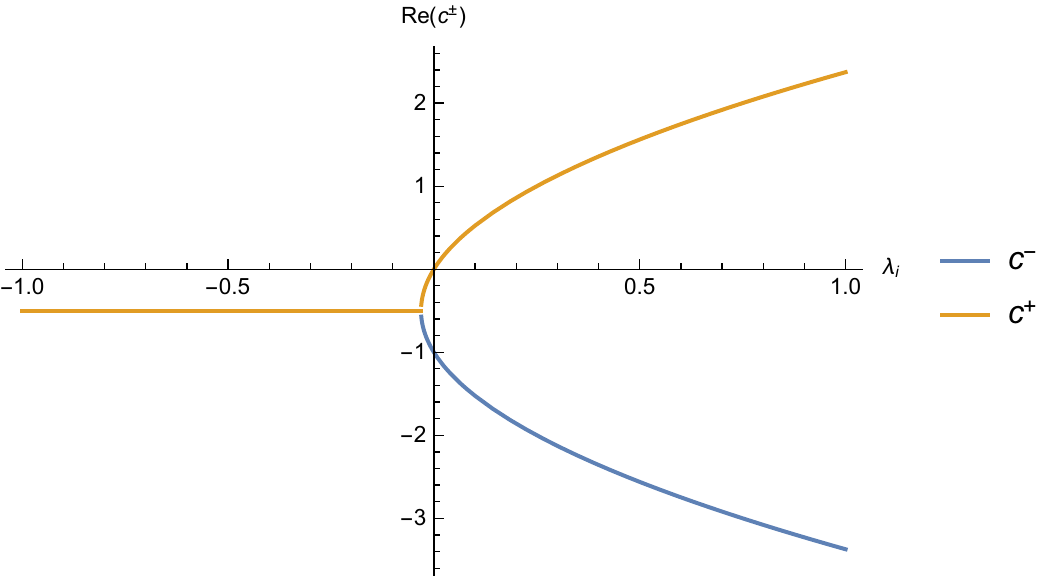}
\caption{\emph{The real part of $c^+$ and $c^-$ vs. $\lambda_i$.}}
\label{Figure_C}
\end{center}
\end{figure}
We can see how if $\lambda$ is negative (as can happen at a saddle point of the shape potential), then the real part of both $c^+$ and $c^-$ is negative. This means that, if~we want to impose that the solution converges as $\tau \to 0$, we will have to set $A^+_i = A^-_i =0$ for each negative~eigenvlue.

If the eigenvalue is positive, then we see from the plot that $c^+>0$ while $c^-<0$ for all $\lambda_i>0$, so we have to set $A^-_i =0$.

\subsection{Generalization to Arbitrary $N$ and Non-Zero~Energy}

To generalize the result from $N=3$ to arbitrary $N$, we consider the kinetic metric on the extended configuration space:
\begin{equation}
\sum_{a=1}^N m_a \, d \bm r_a \cdot d \bm r_a \,, 
\end{equation}
this, in~terms of the mass-rescaled Jacobi coordinates, becomes
\begin{equation}
\sum_{a=1}^{N-1}  d \bm \rho_a \cdot d \bm \rho_a + d \bm r_\st{cm} \cdot d \bm r_\st{cm}  \,
\end{equation}
where ${\bm \rho_a}$ coordinatize the \emph{relative} configuration space (the configuration space quotiented by translations).

Now, we can separate the scale and the scale-invariant degrees of freedom by defining the (square root of the) center-of-mass moment of inertia, the~\emph{scale coordinate}:
\begin{equation}
r = \sqrt{\sum_{a=1}^{N-1}  d \bm \rho_a \cdot d \bm \rho_a} \,,
\end{equation}
and the translation-invariant configuration space appears now as the Cartesian product between the scale coordinate $r \in \mathbbm{R}^+$ and a $(3N-4)$-dimensional hypersphere which we call pre shape space. ({{Pre} shape space is the quotient of the extended configuration space, $\mathbbm{R}^{3N}$, by~dilatations and translations alone (keeping the redundance due to rotations)}.)

To further quotient global rotations out, we need to exploit the fact that they act as an $SO(3)$ subgroup of the rotation group $SO(3N-3)$ that realizes the isometries of the $(3N-4)$-sphere of pre shape space. Quotienting a sphere by a subgroup of its rotation group always results in another sphere. In~our case, the~end result is a $(3N-7)$-sphere: shape space.
The kinetic metric then decomposes according to an analog formula to Chazy's kinetic energy decomposition theorem
and, in~the hyperspherical coordinates on pre-shape space, can be written as
\begin{equation}
\begin{aligned}
dr^2 + d \bm r_\st{cm} \cdot d \bm r_\st{cm} + \frac{1}{r^2} d\varphi_1^2 +  \frac{1}{r^2} \left( \sum_{A=2}^{3N-7} \prod_{B=1}^{A-1} \sin^2 \varphi^B \, {d\varphi^A}^2     +  \sum_{a=1}^{N-1} \|\bm \rho_a \times d \bm \rho_a \|^2 \right)=
\\
= dr^2 +\frac{1}{r^2} \left( g_{AB}(\varphi) d\varphi^A d \varphi^B + d \bm r_\st{cm} \cdot d \bm r_\st{cm}  +  \sum_{a=1}^{N-1} \|\bm \rho_a \times d \bm \rho_a \|^2  \right)\,.
\end{aligned}
\end{equation}

There are $3N-7$ conjugate momenta to $\varphi^A$, $\pi_A$, $A= 1,\dots ,3N-7$, one conjugate momentum to $r$, $p_r$, 3 conjugate momenta to $\bm r_\st{cm}$, $\bm p^\st{cm}$ and  3 components of the total angular momentum $\bm L = \sum_{a=1}^{N-1}\bm \rho_a \times  \bm \pi_a   $. The~kinetic energy can  then be decomposed as
\begin{equation}
\frac 1 2 \sum_{a=1}^N  \frac{\| \bm p^a \|^2}{2m_a} = \frac 1 2 p_r^2 + \frac 1 2 \| \bm p^\st{cm} \|^2 + \frac 1 {2r} \left( \| \bm p^\st{cm} \|^2 +   \sum_{A,B=1}^{3N-7} g^{AB}(\varphi^C) \pi_A \pi_B \right)  \,,
\end{equation}
where $g^{AB}(\varphi^C)$ is the inverse of the hyperspherical metric, and~the Newton potential can be written as
\begin{equation}
V_\st{New}(\bm r_a) = \frac 1 r \, C_\st{S}( \varphi^A ) \,,
\end{equation}
with no dependence on the coordinates of the center of mass, which of course implies that their equations of motion are $\ddot {\bm r}_\st{cm} =0$ and their motion can be decoupled from the rest. Assuming now that the angular momentum is zero, and~after reabsorbing the kinetic energy of the center of mass into $E$, the~Hamiltonian constraint takes the form
\begin{equation}
\frac 1 2 p_r^2  + \frac 1 {2 r^2} \sum_{A,B=1}^{3N-7} g^{AB}(\varphi^C) \pi_A \pi_B - \frac 1 r \, C_\st{S}( \varphi^A ) - E = 0 \,.
\end{equation}

If the total energy $E$ is zero, by~replacing $p_r = \tau /r$, and~solving for $r$, we get a unique~solution:
\begin{equation}
r \, C_\st{S}( \varphi^A )- \frac 1 2 \tau^2   - \frac 1 2\sum_{A,B=1}^{3N-7} g^{AB}(\varphi^C) \pi_A \pi_B   = 0 \,,
\end{equation}
and the corresponding pre-shape space Hamiltonian is:
\begin{equation}
H = \log \left(  \tau^2   +  \sum_{A,B=1}^{3N-7} g^{AB}(\varphi^C) \pi_A \pi_B\right) - \log  C_\st{S}( \varphi^A ) - \log 2 \,.
\end{equation}
\begin{equation}
\begin{aligned}
\frac{d \varphi^A}{d \tau} &= 2 \frac{\sum_{A,B=1}^{3N-7} g^{AB}(\varphi^C) \pi_B }{ \tau^2   +  \sum_{A,B=1}^{3N-7} g^{AB}(\varphi^C) \pi_A \pi_B} \,, 
\\
\frac{d \pi_A}{d \tau} &= - 2 \frac{\sum_{A,B=1}^{3N-7} \frac{\partial g^{CD}(\varphi)}{\partial \varphi^A} \pi_C \pi_D }{ \tau^2   +  \sum_{A,B=1}^{3N-7} g^{AB}(\varphi^C) \pi_A \pi_B}  + \frac 1 {C_\st{S}} \frac{\partial C_\st{S}}{\partial \varphi_A} \,, 
\end{aligned}
\end{equation}
the structure of the equations is identical to those for the $3$-body problem~(\ref{DecoupledSystem}). If~now we expand to first order around $\varphi^A  = \varphi^A_0$ (the coordinates of a central configuration), and~$\pi_A = 0$, we get:
\begin{equation}
\begin{aligned}
\frac{d \delta \varphi^A}{d \tau} &= 2 \frac{\sum_{A,B=1}^{3N-7} g^{AB}(\varphi^C_0) }{ \tau^2} \delta \pi_B  \,, 
\\
\frac{d \delta \pi_A}{d \tau} &=   H_{AB} \delta \varphi^B \,, 
\end{aligned}
\end{equation}
where $H_{AB} = \frac{\partial^2 \log C_\st{S}}{\partial \varphi_A \partial \varphi_B} $ is the Hessian matrix. Now, the~Hessian can be diagonalized as before, but~there are three zero eigenvalues associated to the directions corresponding to global rotations. For~these, we have to put the corresponding momenta to zero, because~they are equal to the three components of the total angular momentum of the system, and, unless~that is zero, the~total collision cannot take place. For~those three pre-shape space degrees of freedom, therefore, the~equations of motion just say that they are constants, and~their conjugate momenta are zero. One is then left with $3N-7$ effective equations, one for each independent true shape degree of freedom, of~the form:
\begin{equation}\label{AsymptoticsTotalCollisionEquations}
\frac{d  \rho_i}{d\tau} \propto \frac{\pi_i}{ \tau^2 } \,,
 \qquad
\frac{d   \pi_i}{d\tau} =  \lambda_i \rho_i \,,
\end{equation}
with $\lambda_i$ real constants depending only on the mass ratios $m_a/m_b$.

If the total energy is not zero, one has a quadratic equation to solve for $r$, but~to avoid having to deal with multiple solutions we can exploit the fact that $r$ is small near the total collision, and~solve the Hamiltonian constraint perturbatively:
\begin{equation}
r \approx \frac{\tau^2 + \sum_{A,B=1}^{3N-7} g^{AB}(\varphi^C) \pi_A \pi_B}{C_\st{S}} -  \frac{E \left( \tau^2 + \sum_{A,B=1}^{3N-7} g^{AB}(\varphi^C) \pi_A \pi_B \right)^2}{C_\st{S} \left[ C_\st{S} + 2 E \left( \tau^2 + \sum_{A,B=1}^{3N-7} g^{AB}(\varphi^C) \pi_A \pi_B \right) \right] } \,,
\end{equation}
so
\begin{equation}
H \approx  H_{E=0} + \log \left[ 1 -  \frac{E \left( \tau^2 + \sum_{A,B=1}^{3N-7} g^{AB}(\varphi^C) \pi_A \pi_B \right)}{ C_\st{S} + 2 E \left( \tau^2 + \sum_{A,B=1}^{3N-7} g^{AB}(\varphi^C) \pi_A \pi_B \right) } \right]  \,.
\end{equation}

The corresponding equations of motion acquire deformation terms which, at~first order in $\pi^A$ and $\varphi^A - \varphi^A_0$, take the form:
\begin{equation}
\begin{aligned}
\frac{d \delta \varphi^A}{d \tau} & \approx \left.  \frac{d \delta \varphi^A}{d \tau} \right|_{E=0} -\frac{2 E g^{AB}(\varphi_0) \pi_B}{C_\st{S}(\varphi_0)}\,, 
\\
\frac{d \delta \pi_A}{d \tau} &\approx   \left. \frac{d \delta \pi_A}{d \tau}\right|_{E=0}  -\frac{E \,  \tau ^2 \frac{\partial^2 C_\st{S}(\varphi_0)}{\partial \varphi^A \partial \varphi^B} \delta \varphi^B}{\left(C_\st{S}(\varphi_0)+\tau ^2 E\right) \left(C_\st{S}(\varphi_0)+2 \tau ^2 E \right)}  \,, 
\end{aligned}
\end{equation}
and both deformation terms are irrelevant as $\tau \to 0$, compared to the undeformed one which diverge like $\tau^{-2}$.

\subsection{The Stratified Manifold of the Total-Collision~Solutions}

Each central configuration comes with $3N-7$ real eigenvalues $\lambda_i$. Depending on the nature of the central configurations, they may all be positive (if we are at the minimum of $C_\st{S}$---the equilateral triangle in the $N=3$ case---which has been conjectured to be unique for all $N$), or~some of them may be negative (in the case of a saddle point, like the three collinear Euler configurations in the three-body problem).

From {Figure}~\ref{Figure_C}, we see that each negative eigenvalue corresponds to a pair of exponents $c^+$, $c^-$ with negative real part, and~therefore the corresponding shape degree of freedom cannot hope to converge to its central-configuration value, unless~both integration constants $A^+$ and $A^-$ are set to zero. On~the other hand, for~each positive eigenvalue, the~integration constant $A^-$ has to be put to zero in order for the solution to~converge.

Let us assume first that, at~the central configuration of interest, there are $M$ \emph{{distinct}} positive $\lambda_i$'s and $3N-7-M$ negative ones, and~assume also that the positive eigenvalues are ordered from smallest to largest, $\lambda_1 < \lambda_2 < \dots < \lambda_M $. Then only $M$ integration constants remain unspecified, and~the solutions are of the following form:
\begin{equation}
\rho_1 = A^+_1 \, \tau^{c^+(\lambda_1)}  \,, ~ \dots \,, \rho_M = A^+_M \, \tau^{c^+(\lambda_M)}  \,, ~~ \rho_{M+1} = 0 \,, ~ \dots  ~ \rho_{3N-7} = 0 
\end{equation}

The above describe a $(M-1)$-parameter family of distinguished solutions, because~any two choices of integration constants, $(A^+_1, \dots A^+_M)$ and $(A'^+_1, \dots A'^+_M)$ that are related by
\begin{equation}
(A'^+_1, \dots A'^+_M) =  (k^{c^+(\lambda_1)}A^+_1, \dots k^{c^+(\lambda_M)} A^+_M) \,,  \qquad k >0 \,,
\end{equation}
describe the same curve in shape space, just parametrized differently. Within~this $(M-1)$-dimensional manifold, there are special regions corresponding to the cases in which certain integration constants are zero. Let us, in~what follows, list all the possible distinct~cases.

\begin{figure}
\begin{center}
\includegraphics[width=0.4\textwidth]{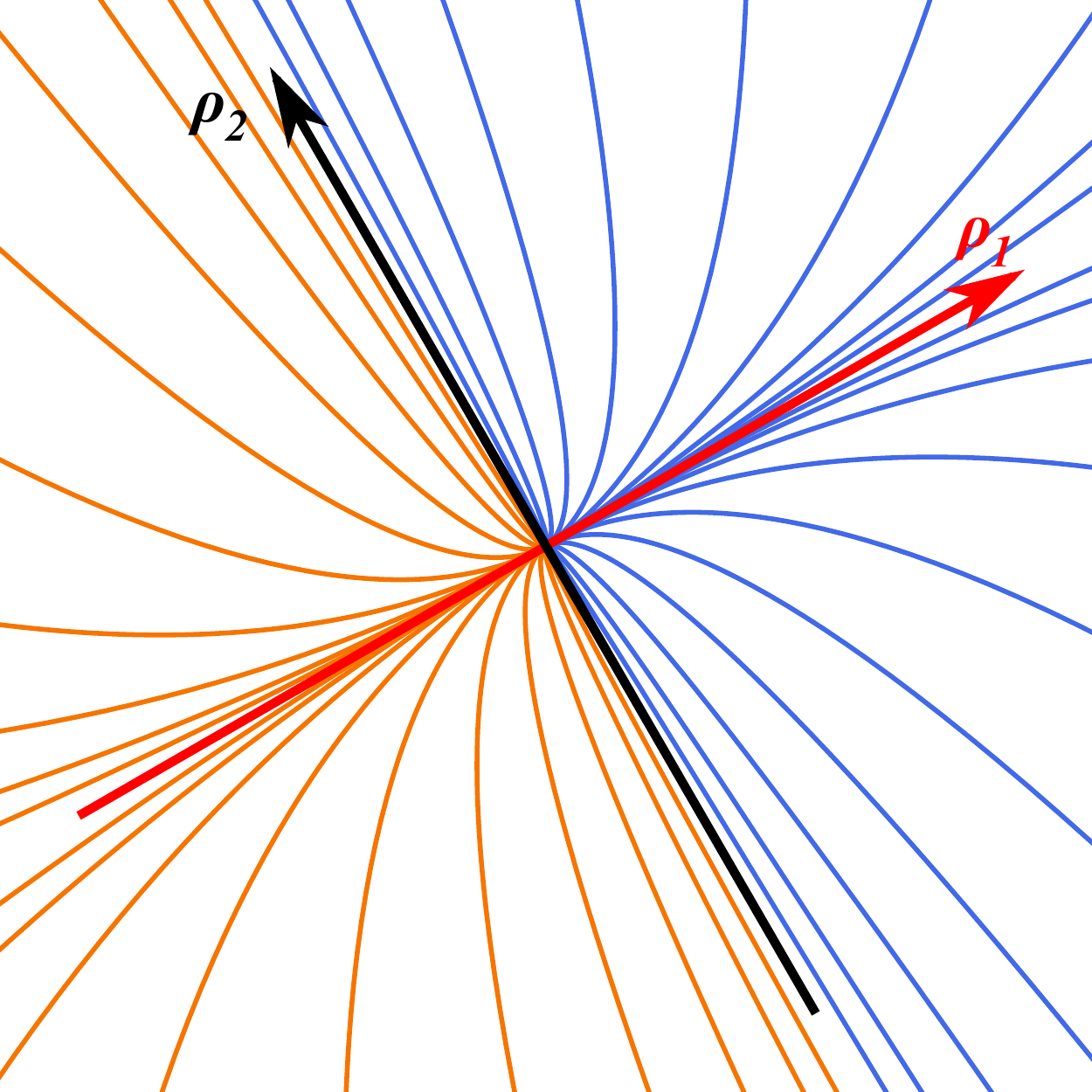}
\caption{\emph{The total-collision solutions of the 3-body problem, when the asymptotic shape is an equilateral~triangle.}}\label{3BodySolFig}
\end{center}
\end{figure}

If $A^+_1 \neq 0$, then the solution curves all approach the central configuration with the same tangent, parallel to the principal eigendirection $\rho_1$ (the one corresponding to the largest eigenvalue), and~away from it they splay out in all $\rho_2,\dots \rho_M$ directions, at~a pace that is determined by the values of the other integration constants $A^+_2,\dots,A^+_M$. \linebreak This is easy to prove: the tangent vector to the parametrized curves is \linebreak $\left( c^+(\lambda_1)  A^+_1 \, \tau^{c^+(\lambda_1)-1} , \dots, c^+(\lambda_M)  A^+_M \, \tau^{c^+(\lambda_M)-1}  \right)$, and~normalizing it to one we get a vector that, in~the limit $\tau \to 0$, tends to $(1,0,\dots,0)$. 
Moreover, these solutions can be divided in two disjoint components, according to whether $A^+_1$ is positive or negative. The~former approach the central configuration along the positive-$\rho_1$ axis, the~latter along the negative one. In~{Figure}~\ref{3BodySolFig}, we show an example of this family of solutions for the 3-body~problem.

\begin{itemize}
\item
If $A^+_1$ is zero, but~$A^+_2 \neq 0$, the~solutions lie in the $\rho_1=0$ subspace, and~the analysis exposed above can be repeated within this subspace, this time with $\lambda_2$ playing the role of principal eigenvalue. The~solutions all approach the central configuration tangentially to the $\rho_2$ axis, and~they divide into two connected components, according to the sign of $A^+_2$;

\item
If $A^+_1 = A^+_2 = \dots = A^+_L = 0$, $L<M$, then the solution lies in the subspace $\rho_1 = \rho_2 = \dots = \rho_L = 0$, and~the role of principal eigendirection is played by $\rho_{L+1}$, and~the solutions are asymptotically tangent to $\rho_{L+1}$, and~belong to two disconnected components, according to the sign of $A^+_{L+1}$;

\item
If only $A^+_M \neq 0$, then there are only two solutions, which remain always on the positive ({respectively} 
negative) $\rho_M$ axis;

\item
Finally, if~all the $A^+_i$ are zero, the~solution is only the homothetic one, which never changes shape as it falls into a total collision.
\end{itemize}

What we just described is a \emph{stratified manifold} of solutions, in~which each stratum is obtained from the higher one as the special case in which the first non-zero integration constant of the stratum above is set to~zero.

In the case of \emph{{degenerate}} eigenspaces (when two or more eigenvalues are identical, which happens for example in the three-body problem when the three masses are equal), the~count of free integration constants does not change, and,~therefore, the dimension of the space of solution is the same as above, as~is its structure as stratified manifold. What changes is the fact that, when the degenerate eigenvalue is the principal one (because it is the smallest, or~because the integration constants associated to the eigendirections of smaller eigenvalues have been all put to zero), the~solution curve can approach the total collision from any direction within the degenerate~eigenspace.

\subsection{The Essential Singularity of Total~Collisions}

In the previous subsection, we have shown how the total-collision solutions can only approach the central configuration along one of the eigendirections of the Hessian matrix that are associated to a positive eigenvalue. Moreover, we have shown that, in~the case of distinct eigenvalues, the~solutions that approach the total collision from the eigendirection associated to the lowest positive eigenvalue are just two. The~ones approaching it from the second-smallest eigendirection are two disjoint one-parameter families; the ones approaching from the third-smallest eigendirection are two disjoint two-parameter families, and~so on, all the way to the highest stratum, which consists of two disjoint $(M-1)$-parameters families of solutions. The~largest possible stratum of solutions for $N$ particles can be obtained in the case in which all $3N-7$ eigenvalues are positive, which means that the corresponding central configuration is a minimum of the complexity function. Then, there is a stratum which is ($3N-8)$-dimensional. So, for~example, in~the unequal-mass three-body problem, if~the total collision asymptotes to an equilateral triangle (the absolute minimum of the complexity function), we get two one-parameter families of~solutions.

We know what the tangent to these solution curves does, but~knowing the tangent is not enough to fix all integration constants $A^+_i$, while the values of the integration constants determine the solution. Since we are interested in investigating the possibility of continuing each solution in a unique way through the total collision, we want to check whether there exist some variables whose values fix all integration constants, and~are well-defined at the total collision. One might look for such `manifest causes' in the geometry of the curve on shape space, which, according to the conjecture at the  basis of shape dynamics, captures all there is to know about physical reality. However, one can show that, in~the generic case ({{that is}, when none of the constants $c^+(\lambda_i)$ are commensurable}), no differential quantity defined on shape space can fix these integration constants, because {at total collisions we have an \emph{essential}  \emph{singularity.}} We can see this in this way. Consider the normalized $n-th$ $\tau$-derivative vector of our solution curve:


\begin{equation}\label{NormalizedNthDerivative}
\begin{aligned}
 \frac{1}{\sqrt{\sum_{i=1}^M (A_i^+)^2  \prod_{k=0}^{n-1} (c^+(\lambda_i) -k)^2 \tau^{2c^+(\lambda_i)  - 2n} }} \cdot \\
\cdot \left( \begin{array}{c} A_1^+  \prod_{k=0}^{n-1} (c^+(\lambda_1) -k) \tau^{c^+(\lambda_1)  - n} \\ \vdots \\  A_M^+ \prod_{k=0}^{n-1} (c^+(\lambda_M) -k) \tau^{c^+(\lambda_M)  - n} \end{array}\right)\,.
\end{aligned}\end{equation}


As $\tau \to 0$, this quantity asymptotes to
\begin{equation}\label{NormalizedNthDerivativeAsymptote}
\text{\it sign} \left( A_1^+  \prod_{k=0}^{n-1} [c^+(\lambda_1) -k] \right) \left(   1 , 0 ,  \dots ,0 \right) \,.
\end{equation}

So, imagine we want to join two curves that asymptote to the same central configuration, characterized by integration constants $A^+_i$ and $A'^+_i$, one reaching the total collision from below ($\tau \to 0^-$) and one from above ($\tau \to 0^+$). They both reach the same point at $\tau =0$, so whatever pair of curves we choose, they will always be continuous. Now, ask that their tangent is continuous: we want the  normalized first derivatives to match. This~imposes
\begin{equation}
\text{\it sign} \left( A_1^+ \right) = - \text{\it sign} \left( A_1'^+ \right) \,,
\end{equation}
that is, the~two curves have to approach $\rho_1=0$ from the two opposite directions. This can be immediately seen from Figure~\ref{3BodySolFig} in the 3-body case.
However, if~now we hope to fix any further relations between integration constants by asking that any further normalized derivative is continuous, we are disappointed. Once we assume that $A_1^+$ and $A_1'^+$ have opposite signs, all derivatives are automatically continuous. We could join any two curves in Figure~\ref{3BodySolFig}, provided they live in opposite sides of the black axis, and~they would always be infinitely differentiable.  This is a behavior that signals the presence of an essential singularity: for example the function $e^{- 1/x}$ at $x\to 0$ tends to zero, as~do all of its derivatives. This function is not analytic in zero, because~it is the inverse of $e^{1/x}$, which is a textbook example of essential singularity (the function and all of its derivatives diverge in zero). 

There are exceptions to this result, in~the exceptional case in which, due to the particular values of the eigenvalues $\lambda_i$ and $\lambda_j$, the~ratio of the associated constants $c^+(\lambda_i)/c^+(\lambda_j)$ is a rational number. Then, in~this case, there exist integers $\alpha$ and $\beta$, such that the variables
$\rho_i^{\alpha}$ and $\rho_j^{\beta}$ admit the finite ratio $(A_i^+)^\alpha/(A_j^+)^\beta$ at $\tau  \to 0$, which allows us to extract some information on the integration constants $A_i^+ $ and $A_j^+$ at the singularity. Then, if~all $M$ positive eigenvalues are such that the corresponding constants $c^+(\lambda_i)$ are commensurable, we can define a set of $M$ variables, by~raising the $\rho_i$ to appropriate integer powers, that all tend to zero at $\tau \to 0$ as the same power of $\tau$. The~simplest such case is that of all-equal eigenvalues, where all $\rho_i$ converge to zero with the same power law. Then, in~this case, all solutions can be continued uniquely at the singularity, and~there is a simple change of variables that makes the equations of motion regular there. These cases, however, account for a countable set of choices of masses, and~the generic situation is that described above, of~an essential singularity preventing~continuation.

\section{Conclusions}\label{sec4}

As shown in~\cite{Barbour:2013,FlavioSDBook}, the~dynamics of the $N$-body problem can be equivalently formulated as a non-autonomous system of ODEs on shape space, reducing the system to its irreducible core of physical degrees of freedom. In~this formulation, as~was shown in~\cite{Reichert2021}, the~total-collision solutions can be characterized neatly as solutions that end at a central configuration with zero dilatational momentum and zero shape momenta. The~question then arises, of~whether these solutions can be regularized in the manner of two-body collisions, or~continued through  the singularity similarly to what was done for cosmological solutions of general relativity in~\cite{Koslowski,Flavio2019,Sloan2019}.
Regardless of whether the system has positive or zero energy, the~asymptotics of the total-collision solutions is universal, and~it is captured by Equations~(\ref{AsymptoticsTotalCollisionEquations}), which are completely determined by the eigenvalues of the Hessian matrix of the (log of) the shape potential at the central configuration. If~the central configuration is a minimum of the shape potential, these eigenvalues are all positive and one has a manifold of total-collision solutions of maximal dimension ($3N-8$), and~for each negative eigenvalue, the~dimension of the total collision manifold decreases by one. The~manifold has the structure of a stratified manifold, each stratum obtained by considering the integration constant that were non-zero in the stratum above, and~setting to zero the one that corresponds to the highest eigenvalue. In~each stratum, the~solution curves will approach the singularity tangentially to the eigendirection corresponding to the highest eigenvalue whose integration constant is~non-zero.

At the singularity, unless~one considers very special choices of masses (e.g.,~all identical), the~dynamical system has an essential singularity, which erases (at least some) information regarding all finite-degree derivatives of the dynamical variables, much like the limit $x \to 0$ of the $e^{-1/x}$ function, whose derivatives are all zero at the singularity. This mirrors what was found in certain homogeneous-but-non-isotropic cosmological models (namely Bianchi IX), where the system, when studied on shape space ({In the case of general relativity, with~shape space we mean the space of conformal 3-geometry. Similarly to what happens in the \emph{N}-body problem, a~curve on this shape space codifies all the information that is necessary to reconstruct uniquely a solution of general relativity~\cite{FlavioSDBook}}) behaves like a chaotic billiard ball (what Misner nicknamed ``mixmaster behavior'') which bounces an infinite amount of times in any finite proper-time interval ending at the big bang singularity. This, too, is an essential singularity: the limit set of the dynamics is the border of shape space (which has the topology of a circle), but~the location on this border does not admit a well-defined limit, much like the value of $\sin(1/x)$ when $x \to 0$ (another classic example of essential singularity).

In the case of Bianchi IX, however, there is a simple extension of the model that removes this singularity: adding a scalar field whose potential does not grow too fast for large values of the field~\cite{Koslowski,Flavio2019}. The~scalar field changes the asymptotics of the shape momenta in such a way that the ``mixmaster'' chaotic behavior stops after a finite number of bounces, and~the system settles on a so-called ``quiescent''  solution that admits a well-defined limit at the singularity. This is the foundation of the result~\cite{Koslowski,Flavio2019} on the continuation of these solutions through the singularity. Interestingly, this regularization could be attributed to quantum effects, because~the Starobinski potential satisfies the conditions specified in~\cite{Flavio2019} for the onset of quiescence. In~fact, a~scalar field with this particular potential emerges as the lowest-order quantum correction to the Einstein--Hilbert action in an effective field theory approach (it is due to an $R^2$ term in the action).

It is possible that the total collisions of the \emph{N}-body model we studied in the present paper might admit a similar regularization, at~the cost of adding some correction terms to the dynamics, which become relevant only near a singularity. This would, however, be a departure from the purely Newtonian \emph{N}-body problem, and~is beyond the scope of the present~paper.\\

\noindent\textbf{Acknowledgements}\\

\noindent F.M. and P.R. thank Julian Barbour and Tim Koslowski for valuable comments. F.M. thanks the Action CA18108 QG-MM from the European Cooperation in Science and Technology (COST), the Foundational Questions Institute (FQXi) and the Spanish Ministry of Education.

\end{document}